\documentclass[12pt]{elsart}

\usepackage{graphicx}
\usepackage{subfigure}
\usepackage{amsmath}
\usepackage{cite}
\usepackage{endfloat}

\journal{Physics Letters A}

\begin{document}

\begin{frontmatter}

\title{Low-temperature metastable states in a stacked triangular Ising antiferromagnet}
\author{M. \v{Z}ukovi\v{c}\corauthref{cor}},
\ead{milan.zukovic@upjs.sk}
\author{L. Mi\v{z}i\v{s}in},
\author{A. Bob\'ak}
\address{Department of Theoretical Physics and Astrophysics, Faculty of Science,\\ 
P. J. \v{S}af\'arik University, Park Angelinum 9, 041 54 Ko\v{s}ice, Slovak Republic}
\corauth[cor]{Corresponding author.}

\begin{abstract}
We study low-temperature magnetization processes in a stacked triangular Ising antiferromagnet by Monte Carlo simulations. In increasing and decreasing magnetic fields we observe multiple steps and hysteresis corresponding to formation of different metastable states. Besides the equidistant threefold splitting of the 1/3 ferrimagnetic plateau, we additionally confirm a fourth plateau in the field-increasing branch and a sizable remanence when the field is decreased to zero. The newly observed plateau only appears at sufficiently low temperature and sufficiently large exchange interaction in the stacking direction. These observations reasonably reproduce low-temperatures measurements on the spin-chain compound $\rm{Ca}_3\rm{Co}_2\rm{O}_6$. 
\end{abstract}


\end{frontmatter}

\section{Introduction}
\hspace*{5mm} Geometrically frustrated spin systems continue to attract attention owing to their often peculiar or unexpected behavior. This is particularly the case at low temperatures where the geometrical frustration, caused by incompatibility between the lattice geometry and local interactions, results in a high degeneracy of states. One of the simplest geometrically frustrated spin models, with a long history of investigation, is an Ising antiferromagnet on a stacked triangular lattice (IASTL) \cite{berker,blank,copper,hein,kim,netz1,netz2,netz3,netz4,plum1,bunker,diep,nagai1,nagai2,nagai3,plum2,plum3,kurata}. Its importance also stems from the fact that it reasonably describes some real magnetic materials, such as the spin-chain compounds $\rm{Cs}\rm{Co}\rm{X}_3$ (X is Cl or Br) and $\rm{Ca}_3\rm{Co}_2\rm{O}_6$. The latter consists of one-dimensional ferromagnetic spin-chains aligned along the {\it c} axis that form a triangular lattice in the {\it ab} plane. The interchain interaction is antiferromagnetic and is much weaker than the ferromagnetic coupling within the chains. Recently, there has been considerable effort to explain peculiar phenomena observed in magnetization processes of this compound \cite{kage1,kage2,maig,hardy1}. Namely, the magnetization curves as a function of an external magnetic field displayed at low temperatures a significant out-of-equilibrium nature accompanied with a strong hysteresis and splitting of the broad 1/3 magnetization plateau in the ferrimagnetic state into multiple steps. Additionally, different relaxation mechanisms were observed at very low and higher temperatures, with an intermediate regime in between when the two relaxation processes had a comparable influence. In order to interpret the non-equilibrium dynamics, Kudasov et al. \cite{kuda1,kuda2,kuda3,kuda4,kuda5,kuda6} used an analytical approximation and numerical simulations and at least qualitatively explained the appearance of the three steps in the magnetization curves at low temperatures below the saturated state as well as their dependence on a magnetic-field sweep rate and temperature. This intriguing behavior of the magnetization curves was to a large extent also reproduced in the Monte Carlo (MC) simulations \cite{yao1,yao2,yao3,qin,soto} and ascribed to formation of metastable states featuring interlinked mobile domain walls structure \cite{soto}. On the other hand, the mean-field approach was shown to be inadequate to explain such a behavior \cite{yao2}. \\
\hspace*{5mm} Despite the relatively large number of investigations of the phenomena that occur at moderately low temperatures, to our best knowledge, yet no satisfactory explanation has been provided for the behavior observed at very low temperatures. In particular, the experimental studies \cite{maig,hardy1} have shown that at sufficiently low temperatures ($T$ = 2K) the system displays even more metastable states as a function of the applied field. Namely, compared with the moderately low temperature region, the magnetization curve displays not three but four steps before the fully saturated value is reached and the saturation occurs at the critical field which is shifted to a larger value. These features were only observed in the field-increasing (FI) curve, while in the field-decreasing (FD) curve only three steps were detected and the fields at which they appeared were shifted towards lower values. Furthermore, the FD process lead to a considerable remanence. Therefore, the FI and FD branches were shown to be totally irreversible, forming a large hysteresis loop \cite{maig}. Some simple proposal for the magnetic substructures corresponding to different steps were presented in Ref.~\cite{maig}, however, no investigations of this kind have been done so far. \section{Model and simulation}
\hspace*{5mm} In effort to explain the above peculiar features observed in the spin-chain compound $\rm{Ca}_3\rm{Co}_2\rm{O}_6$, we have used MC simulations to study the magnetization processes in the IASTL model, focusing on the low-temperature region. In the present paper we consider the model described by the Hamiltonian 
\begin{equation}
\label{Hamiltonian}
H=-J_{1}\sum_{\langle i,j \rangle}s_{i}s_{j}-J_{2}\sum_{\langle i,k \rangle}s_{i} s_{k}-h\sum_{i}s_{i}\ ,
\end{equation}
where $s_{i}=\pm1$ is an Ising spin, $\langle i,j \rangle$ and $\langle i,k \rangle$ denote the sum over nearest neighbors in the triangular plane and in adjacent planes, respectively, and $h$ is an external magnetic field. The exchange interaction parameters are considered $J_1<0$ and $J_2>0$, which means that the antiferromagnetic triangular planes are coupled ferromagnetically in the stacking direction. Simulated spin systems are of the size $L^3$ with the periodic boundary conditions. We checked several lattice sizes and found that above a certain value the magnetization curves do not change considerably with $L$ and, therefore, we use a moderate size of $L=30$ throughout the paper. The updating follows the Metropolis dynamics and for thermal averaging we typically consider $N=10^4,10^5$ and $10^6$ MCS (Monte Carlo sweeps or steps per spin) after discarding another $N_{0} = 0.2 \times N$ MCS for thermalization. The magnetization versus magnetic field curves are evaluated at a fixed temperature $t=k_BT/|J_1|$ for two cases: when the field increases (FI) from zero to higher values until the magnetization is fully saturated and when the field decreases (FD) back to zero value. For FI (FD) magnetization processes the simulation starts from random (ferromagnetic) initial state and the simulation at the next field value starts from the final state obtained at the previous field value. The magnetization $m$ is evaluated from the equilibrium spin configurations by taking thermal average and normalizing per number of sites. Thus, the saturation value of $m_{sat}=1$ is achieved when all the spins are fully aligned in the field direction.

\section{Results and discussion}
\begin{figure}[t]
\centering
    \subfigure{\includegraphics[scale=0.6]{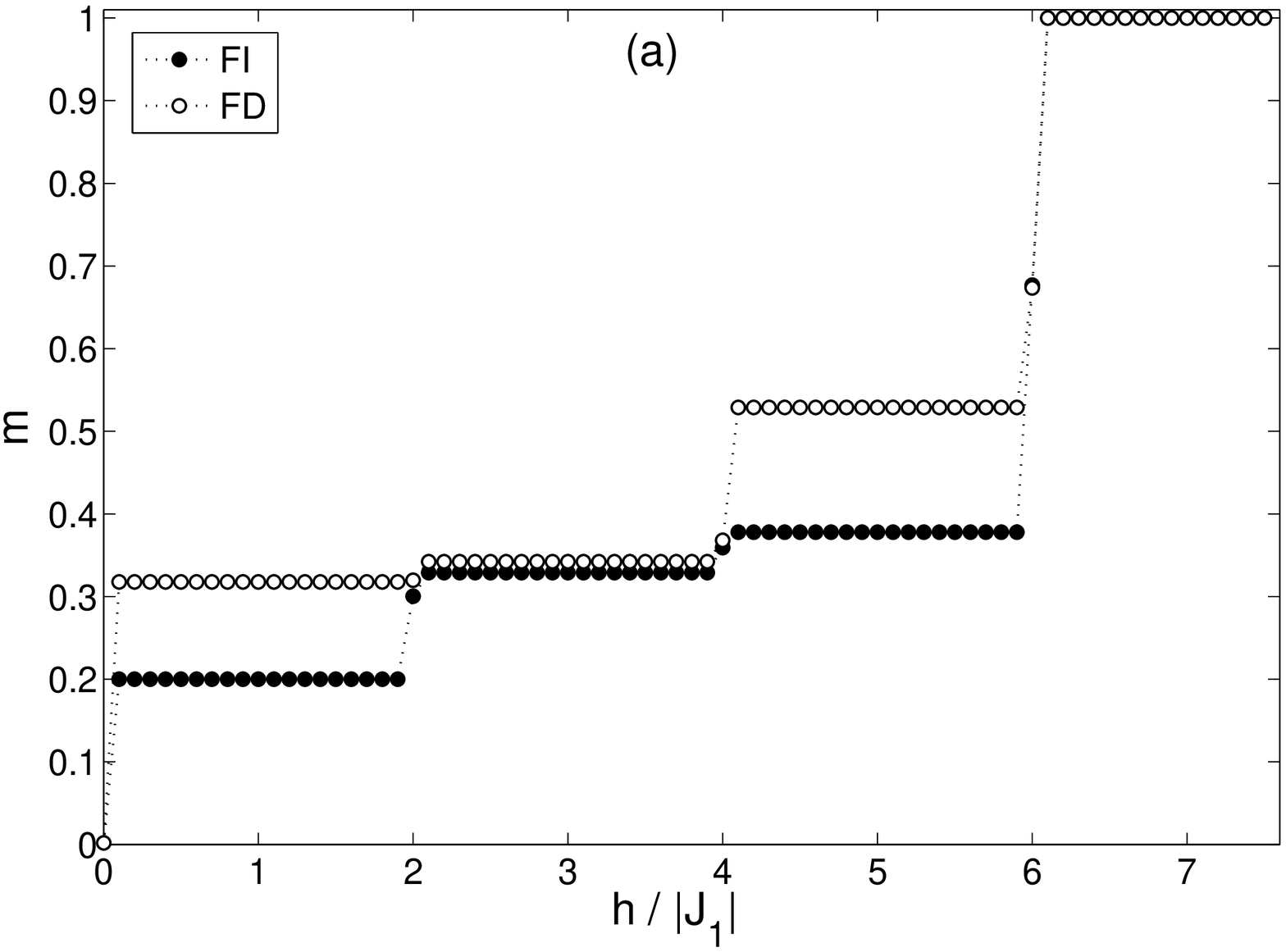}\label{fig:FI-FD_T03}}
    \subfigure{\includegraphics[scale=0.6]{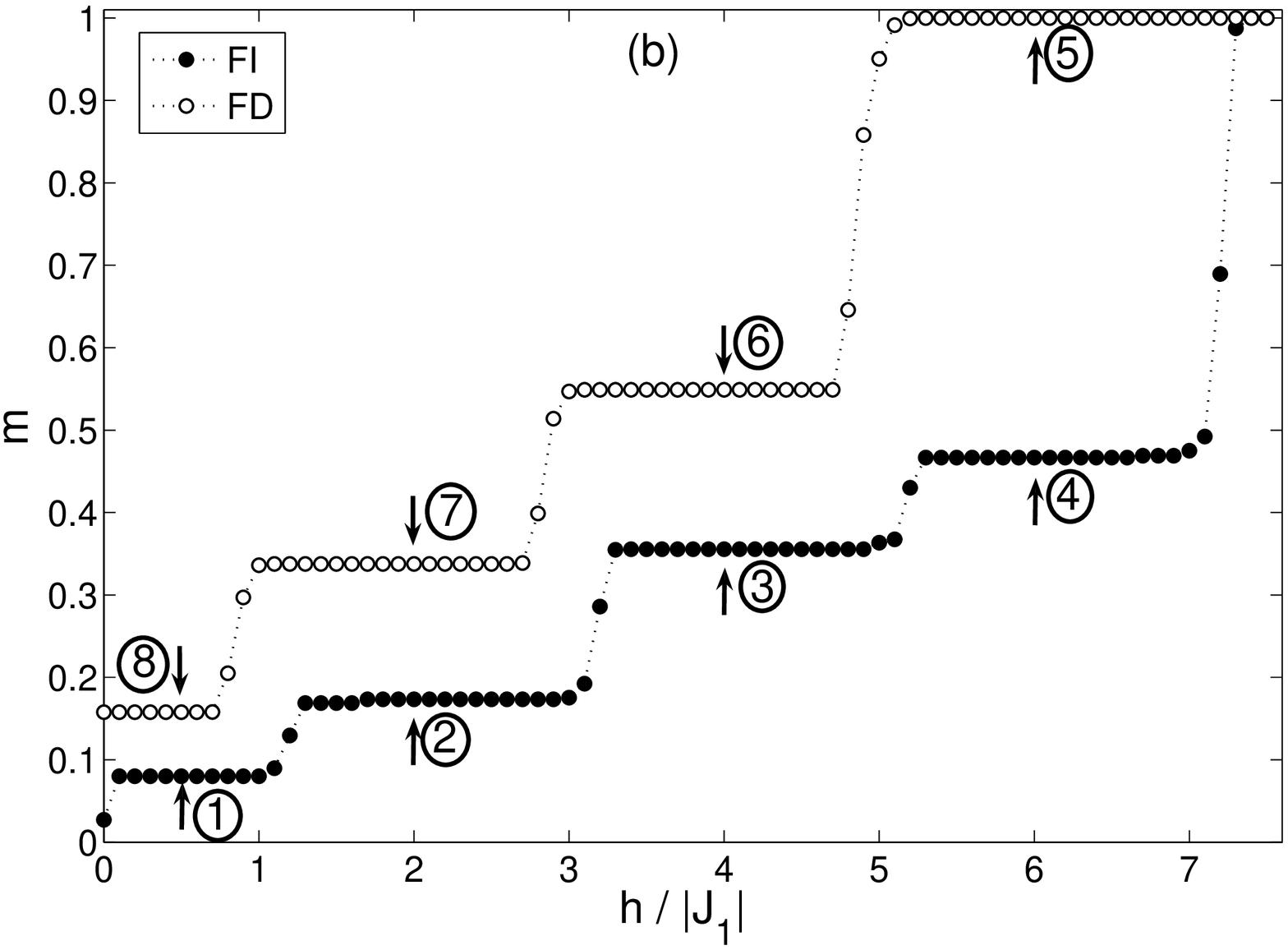}\label{fig:FI-FD_T01}}
\caption{FI and FD magnetization curves for (a) $t=0.3$ and (b) $t=0.1$, with ${\rm MCS}=10^6$ and $J_2/|J_1|=1.$ The numbers in (b) represent different steps and the arrows show the locations where the snapshots were taken.}\label{fig:FI-FD_T0103}
\end{figure}

\begin{figure}[t]
\centering
    \subfigure{\includegraphics[scale=0.6]{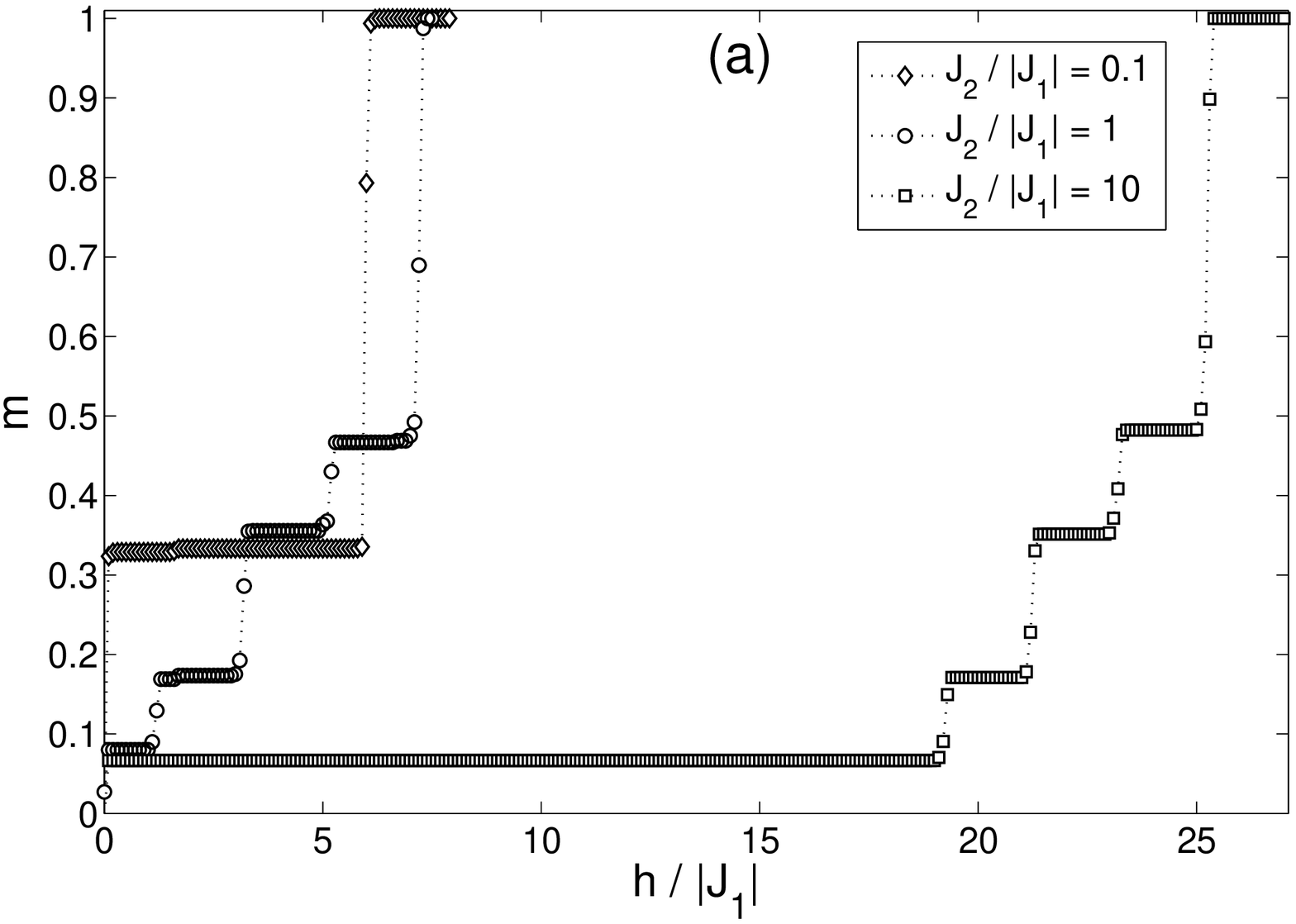}\label{fig:FI-J2_T01}}
    \subfigure{\includegraphics[scale=0.6]{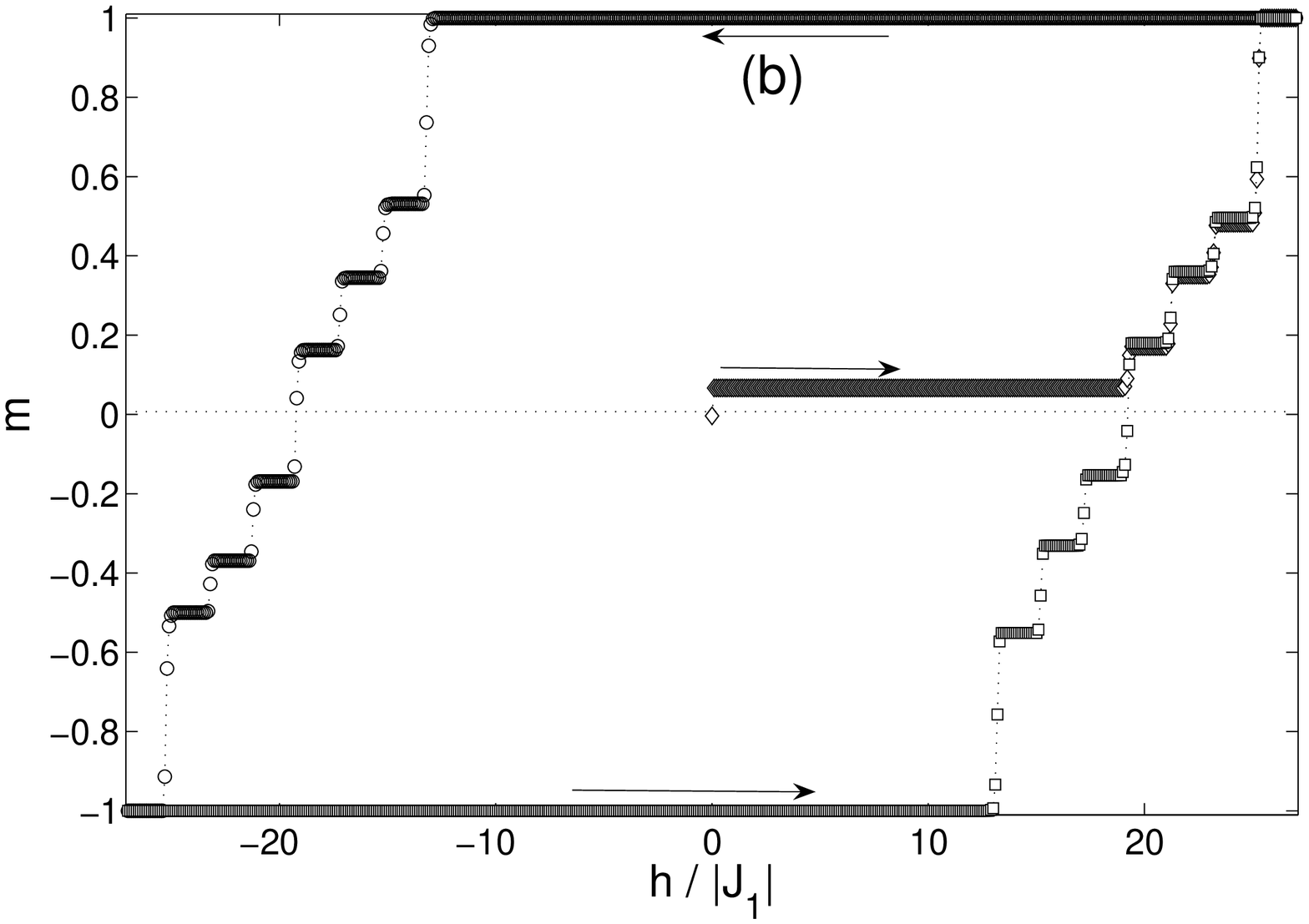}\label{fig:FI-loop_T01}}
\caption{(a) FI magnetization curves for different values of $J_2/|J_1|$=0.1, 1, and 10. (b) FI-FD-FI hysteresis loop for $J_2/|J_1|=10$. The simulations are performed at $t=0.1$ using ${\rm MCS}=10^5$.}
\end{figure}

\begin{figure}[t]
\centering
    \includegraphics[scale=0.6]{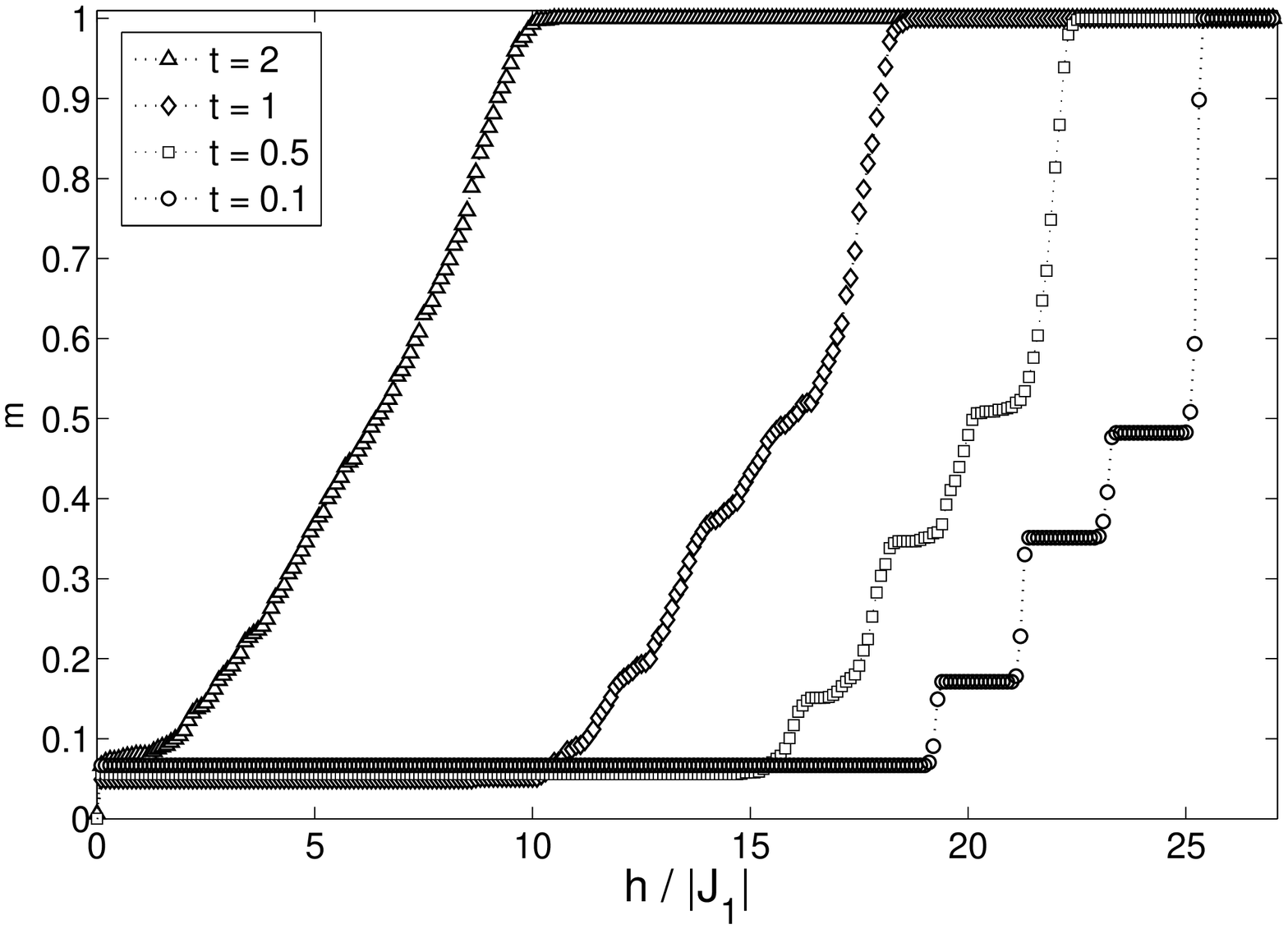}
\caption{FI magnetization curves for different values of $t=0.1$, 0.5, 1, and 2, $J_2/|J_1|=10$ and ${\rm MCS}=10^5$.}\label{fig:FI-T}
\end{figure}

\hspace*{5mm} For moderately low temperatures our MC simulation results corroborate those obtained from other MC simulation studies~\cite{yao1,yao2,yao3,qin,soto} as well as the experiments~\cite{maig,hardy1}. Namely, before the magnetization vs. field curve reaches the saturation value $m_{sat}=1$ at $h_{sat}/|J_1|=6$ in the FI process it displays three equidistant metastable steps. As the field is decreased from high values the geometrical frustration leads to evolution of spin configurations different from that occurred in the FI process, which is manifested by the irreversible character of the FI and FD curves, as shown in Fig.~\ref{fig:FI-FD_T03} for $t=0.3$. There is no remanence and, in accordance with the earlier MC~\cite{yao1,yao2,soto} and experimental~\cite{maig,hardy1} results, with increasing temperature and the number of MCS both steplike branches tend to merge to one broad $m=1/3$ plateau in the ferrimagnetic phase within $0<h/|J_1|<6$ (not shown). \\
\hspace*{5mm} In the following we focus on the behavior displayed at yet lower temperatures. In Fig.~\ref{fig:FI-FD_T01} we present a similar FI-FD magnetization loop obtained at $t=0.1$. A qualitatively different behavior is apparent: the FI branch features not three but four steps below the saturated state and $h_{sat}/|J_1|$ is shifted to a larger value ($\approx 7$). On the other hand, the FD branch retains the three-step character and $h_{sat}/|J_1|$ is shifted to a lower value ($\approx 5$). The last step of the FD branch is shorter and when the field is decreased to zero the magnetization does not vanish. Moreover, the relative magnitude of the remanent magnetization $m_{rem} \approx 0.16$ agrees surprisingly well with that obtained from the measurements on the spin-chain compound $\rm{Ca}_3\rm{Co}_2\rm{O}_6$ at $T=2$K ($\approx 17\%$ of the saturation value)~\cite{maig}. \\
\begin{figure}[t]
\centering
    \subfigure{\includegraphics[scale=0.38]{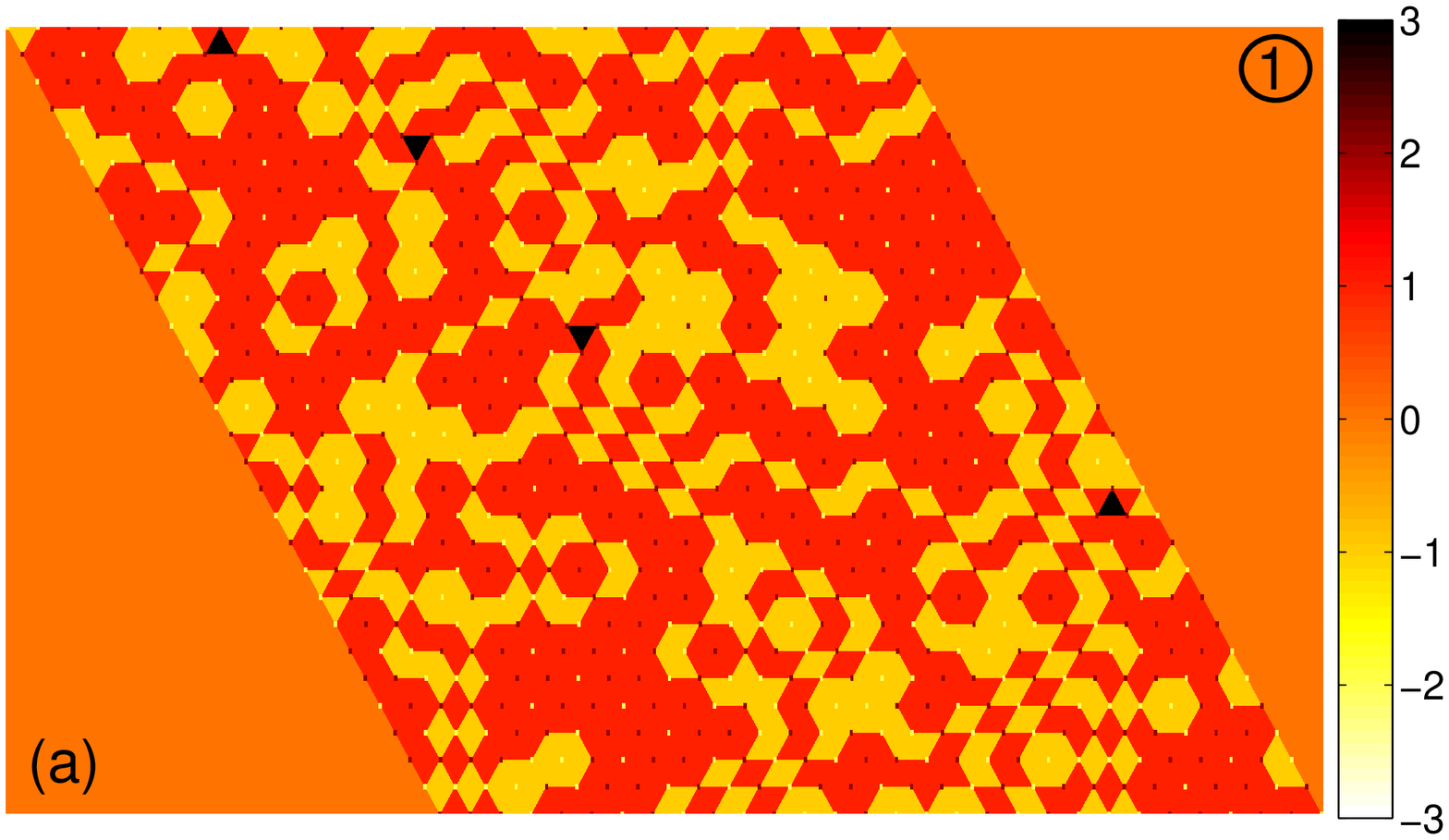}\label{fig:FI_T01_stp1}}
		\subfigure{\includegraphics[scale=0.38]{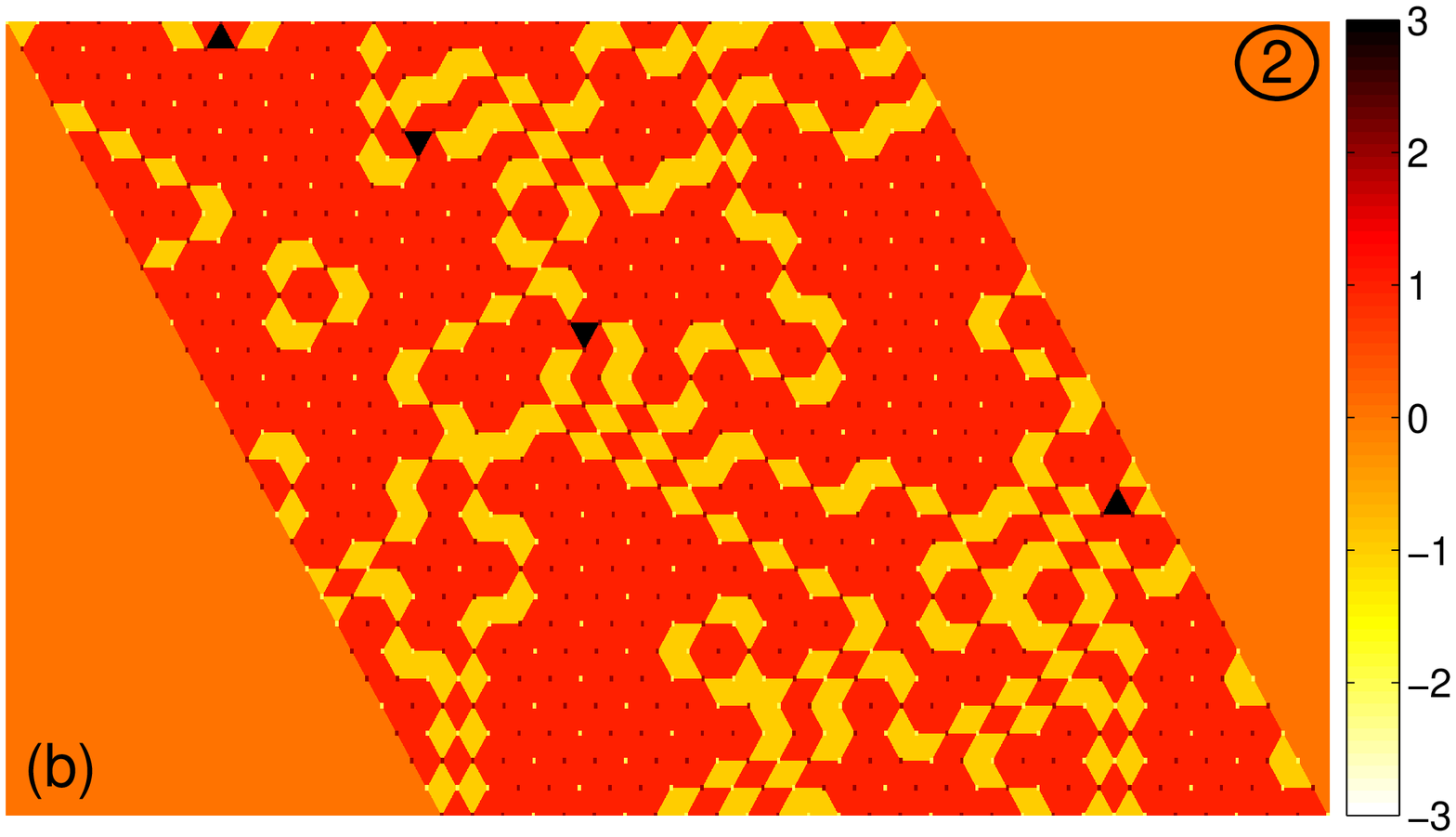}\label{fig:FI_T01_stp2}}
		\subfigure{\includegraphics[scale=0.38]{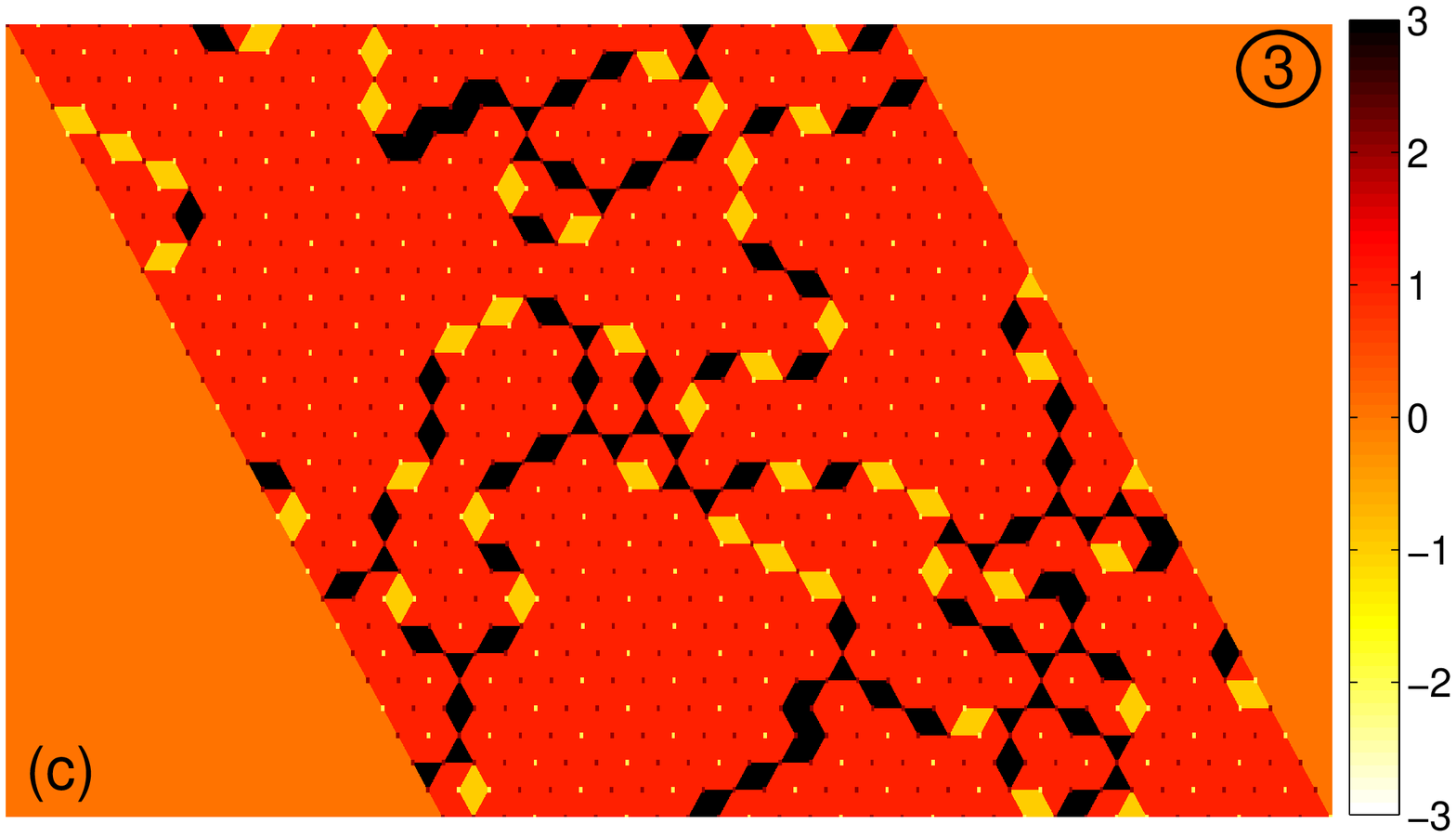}\label{fig:FI_T01_stp3}}
		\subfigure{\includegraphics[scale=0.38]{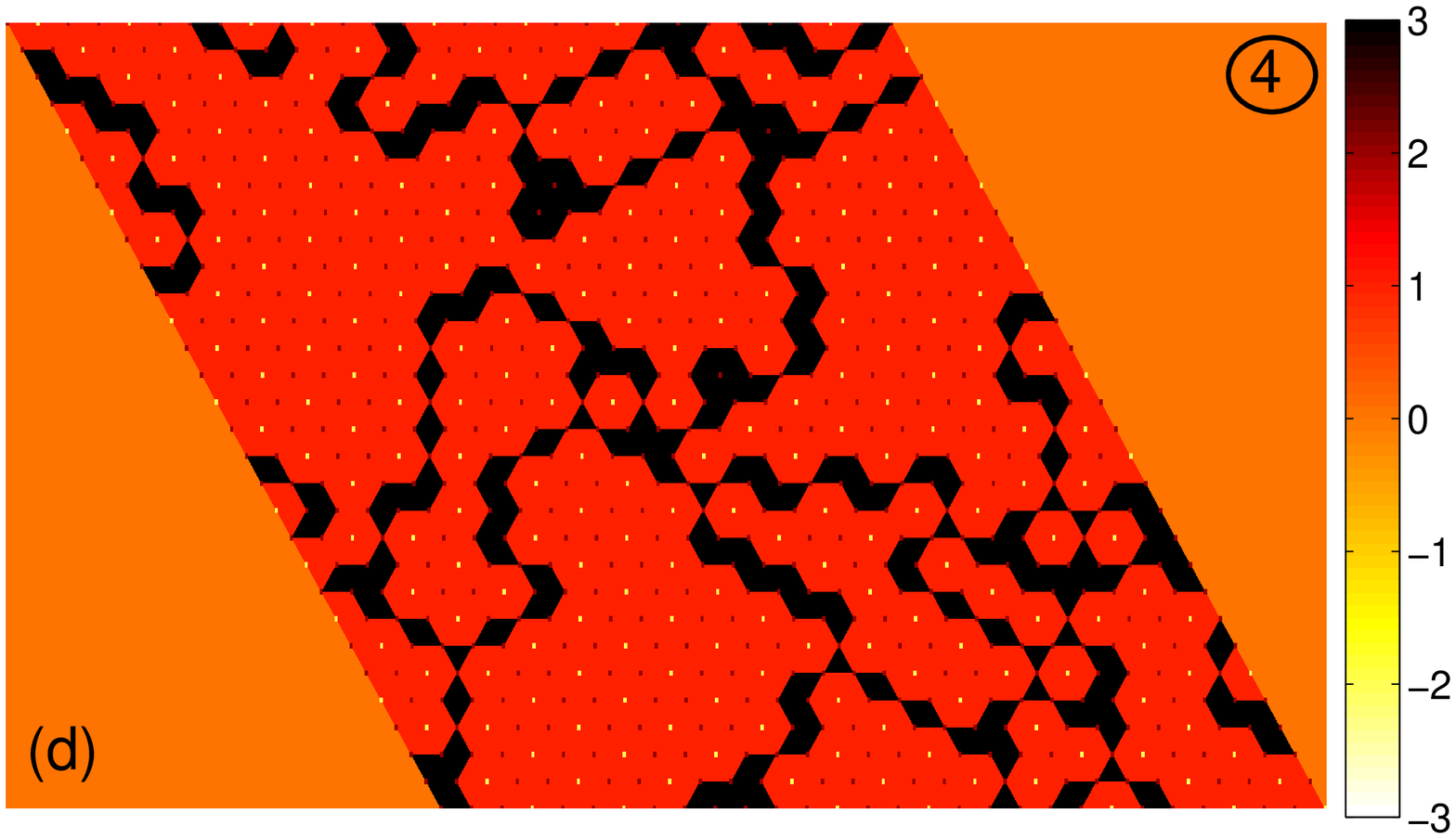}\label{fig:FI_T01_stp4}}
\caption{(Color online) Snapshots of the spin configurations at various plateaus in the FI process corresponding to the field values (a) $h/|J_1|=0.5$, (b) $2$, (c) $4$, and (d) $6$, for $t=0.1$ and ${\rm MCS}=10^5$.}\label{fig:FI_T01_stps}
\end{figure}
\hspace*{5mm} In order to visualize the spin configurations corresponding to the respective steps, we take snapshots during the FI process at different field values of $h/|J_1|=0.5$, 2, 4 and 6 (Fig.~\ref{fig:FI_T01_stps}). The snapshots are taken in a selected plane from the central part of the stack, however, as we show below, except for the low fields they are identical in each plane. Motivated by some previous studies~\cite{maig,soto}, instead of individual spins we visualize elementary triangular plaquettes formed by nearest-neighbor spins. Four different colors (gray scales) correspond to different sums of the spin values on each triangle, as follows: $-3$ - white, $-1$ - yellow (light), $+1$ - red (gray), and $+3$ - black. The snapshots in subfigures (b)-(d) agree well with those obtained by Soto et al.~\cite{soto}, which were interpreted in terms of formation of interlinked mobile domain walls, related to the three steps in the FI magnetization curve. The snapshot of the new state formed at low fields (step 1), shown in subfigure (a), looks similar to the one in subfigure (b) taken at $h/|J_1|=2$ (step 2). However, compared with subfigure (b), it features some scattered yellow (light colored) hexagonal configurations, consisting of six clustered elementary triangles with one (central) spin pointing down and six surrounding nearest neighbors pointing alternatively up and down. Apparently, such configurations are not stable at any finite fields and in step 2 such triangles only appear in the one-dimensional domain walls. Nevertheless, in step 1 the small hexagonal clusters of such triangles persist up to a certain field value ($h/|J_1| \approx 1.2$), thus the total magnetization is reduced in comparison to that observed in step 2. The presence of such configurations at low fields and temperatures (note that such states only appear at sufficiently low temperatures, as already demonstrated in Fig.~\ref{fig:FI-FD_T0103}) can be attributed to the freezing of the entire ferromagnetically ordered chains, comprising the central spins of the hexagons, in the direction opposite to the field. In the chain direction there is no frustration and the ferromagnetic order within the chains can be established already in zero field, however, due to the interchain frustration, the chains remain (partially) disordered \cite{berker,blank,copper,hein,kim,netz1}. When the external field is applied the chains start to order in the ferrimagnetic fashion, nevertheless, the sluggish dynamics at very low temperatures prevents fast rearranging of the entire chains that are pointing in the direction opposite to the field. Only at a sufficiently strong field the Zeeman term overcomes the intrachain exchange energy and the spins in such chains can break the ferromagnetic bonds and flip into the field direction. In the ground state this happens at $h/|J_1|=2J_2/|J_1|$ but thermal excitations at nonzero temperatures tend to shift this field to the lower values until step 1 entirely disappears (see Fig.\ref{fig:FI-FD_T0103}). On the other hand, in the FD branch, below the extended saturation phase (step 5), one can observe only three steps (in Fig.\ref{fig:FI-FD_T0103} denoted by 6, 7 and 8). The snapshots (not shown) reveal patterns similar to those seen in steps 4, 3, and 2, respectively, even though the domains tend to be smaller and the network of the domain walls denser. \\
\begin{figure}[t]
\centering
    \subfigure{\includegraphics[scale=0.6]{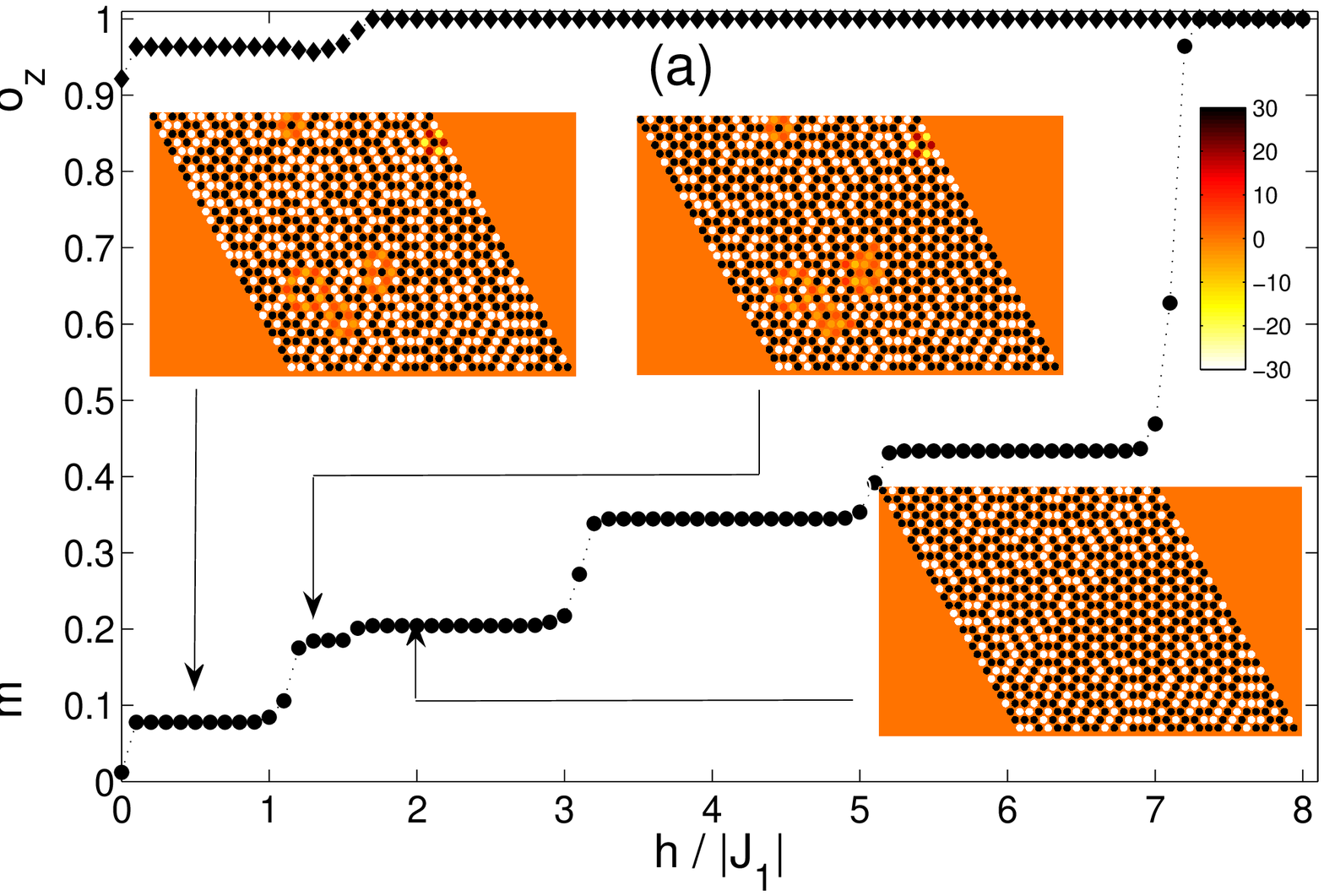}\label{fig:FI_MCS}}
    \subfigure{\includegraphics[scale=0.6]{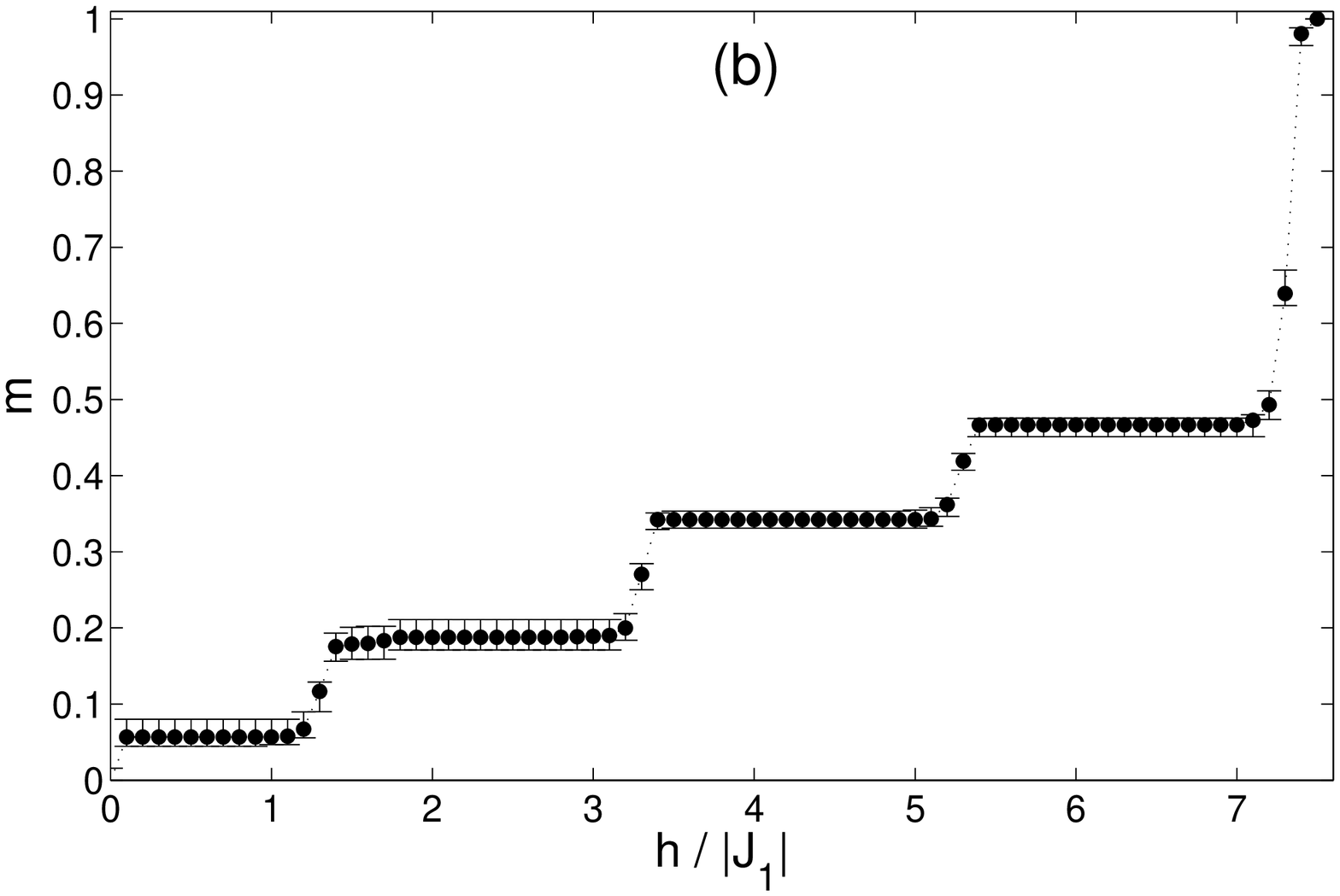}\label{fig:FI_CNF}}
\caption{(a) (Color online) Magnetization $m$ and chain-order parameter $o_z$ of the FI process, for $t=0.1$, $J_2/|J_1|$=1 and ${\rm MCS}=10^6$. The snapshots of the chain magnetizations are taken at $h/|J_1|=0.5$, 1.3 and 2. (b) Magnetization $m$ of the FI process obtained from 10 different simulation runs, with the symbols representing median and the error bars maximum and minimum values. The simulations are performed for $t=0.1$, $J_2/|J_1|$=1 and ${\rm MCS}=10^5$.}\label{fig:FI_STAB}
\end{figure}
\hspace*{5mm} A closer inspection of the FI branch reveals that between steps 1 and 2 (or within step 2) there is an additional smaller step. This step is related to rearrangement of a portion of the spin chains which remain partially disordered throughout step 1 and partially also step 2. This is apparent looking at the order parameter which measures spin ordering within the chains $o_z = \sum_{m=1}^{L^2}\big|\sum_{n=1}^{L}s_{m,n}\big|/L^3$, where the inner summation runs over all spins in the $m$-th chain and the outer over all chains. The snapshots shown in Fig.~\ref{fig:FI_MCS} illustrate the situation now looking at magnetizations of the individual chains. The black and white circles represent spin chains fully ordered in and against the field direction, respectively, while the small domains shown in orange (gray) color represent the chains lacking the full alignment in either direction. Interestingly, within this small step the increasing field initially introduces even more disorder into some chains that were fully aligned against the field direction and located in the neighborhood of the domains of the partially ordered chains. More specifically, the increasing field does not flip these chains as a whole but in fractions, which is manifested in a temporal decrease of the chain-order parameter $o_z$ and small increase of the orange (gray) domains in the snapshot taken at $h/|J_1|=1.3$.\\
\hspace*{5mm} We note that even though the four steps below the saturation phase in the FI branch are expected to merge to one broad plateau of $m=1/3$ when the equilibrium is reached, as shown above, at $t=0.1$ all of them are rather stable and persist up to at least ${\rm MCS}=10^6$. Furthermore, in order to check that the newly observed steps are not produced coincidentally as a result of a particular initialization, we ran 10 simulations starting from different initial states and the steps appeared in each of the obtained magnetization curves. The median values along with the maxima and minima as error bars are presented in Fig. \ref{fig:FI_CNF}. It is worth mentioning that the height of the first step $m \in (0.039m_{sat},0.079m_{sat})$ is comparable with the experimentally identified one at $T=2$K ($m \approx 0.083m_{sat}$)~\cite{maig}. The smaller step connected with the intrachain ordering gets somewhat smeared due to the fact that its position varies to certain extent in different simulations. The effect of using smaller number of ${\rm MCS}=10^5$ is merely reflected in a slight shifting of the critical fields at which the steps occur due to extending of the width of step 1, while steps 2-4 retain the equidistant character. \\
\hspace*{5mm} In Fig. \ref{fig:FI-J2_T01} the FI magnetization curve is shown for different values of the exchange interaction ratio $J_2/|J_1|$. While for the case of $J_2/|J_1|=0.1$ step 1 does not appear at all, since the weak intrachain coupling is easily overcome by a small field, for the case of $J_2/|J_1|=10$ the width of the first step is close to the value $h/|J_1|=2J_2/|J_1|=20$, expected at zero temperature. Thus performing one FI-FD-FI cycle within the two saturated states with $m_{sat}=\pm 1$ a huge hysteresis is formed (Fig. \ref{fig:FI-loop_T01}). Considering the dependence of the width of step 1 on $J_2/|J_1|$, by comparison with the experimental measurements in Ref.~\cite{maig} we can roughly estimate the value of the ferromagnetic exchange interaction in the compound $\rm{Ca}_3\rm{Co}_2\rm{O}_6$. Assuming that the above relation $J_2=h/2$, predicted for $T=0$ K, can reasonably approximate the behavior also at $T=2$ K, from the width of the first step $B \approx 1$ T we can estimate $J_2 \approx 2 \times 10^{-23}$ J. We note that this estimate is rough and the value should be rather viewed as a lower bound, since at the finite temperature of the actual measurement ($T=2$ K) the width of step 1 is expected to be shorter compared to $T=0$ K. This is evidenced from Fig.~\ref{fig:FI-T}, in which we manifest how the FI branch evolves when the temperature is gradually increased. In line with our expectations, due to increasing thermal excitations the first step shrinks and the remaining ones get rounded before they disappear at sufficiently high temperatures. We note that the curves in Fig.~\ref{fig:FI-T} seemingly do not tend to the 1/3 ferrimagnetic plateau with the increasing temperature, as it was the case in our study for $J_2/|J_1|=1$, as well as some previous MC investigations~\cite{yao1,yao2,soto} and the experimental measurements~\cite{maig,hardy1}. In fact, the curves do approach the 1/3 ferrimagnetic plateau, however, the relatively large intrachain coupling makes the dynamics much slower than for $J_2/|J_1|=1$ and, therefore, much more time is required. For example, in our tests at $t=0.5$ it took about $5 \times 10^6$ MCS to reach the 1/3 plateau for $J_2/|J_1|=10$, while one order less MCS was enough for $J_2/|J_1|=1$. Such a tendency can also be observed in Fig.~\ref{fig:FI-J2_T01} in which the curve with the relatively smallest intrachain coupling ($J_2/|J_1|=0.1$) is close to the 1/3 ferrimagnetic state even at very low temperature ($t=0.1$) and using as few as $10^5$ MCS.
  

\section{Conclusions}
\hspace*{5mm} In summary, we studied the low-temperature magnetization processes in the IASTL model using increasing and decreasing magnetic fields. Besides the three metastable steps formed within the 1/3 ferrimagnetic plateau, already observed in some previous studies, we found an additional forth plateau in the field-increasing branch and the remanence in the field-decreasing branch. The critical field values at which the plateaus occurred were shifted in the field-increasing process to the higher and in the field-decreasing process to the lower values, thus displaying a considerable hysteresis effect. The newly observed plateau only showed at sufficiently low temperature and large enough values of the intrachain and interchain interaction ratio, and its width considerably depended on these two parameters, besides some weaker dependence on the number of MCS. The results obtained by the current simple model are in a fairly good agreement with the experimental findings in the low-temperature behavior of the spin-chain compound $\rm{Ca}_3\rm{Co}_2\rm{O}_6$.

\section*{Acknowledgments}
This work was supported by the Scientific Grant Agency of Ministry of Education of Slovak Republic (Grant No. 1/0234/12). The authors acknowledge the financial support by the ERDF EU (European Union European regional development fund) grant provided under the contract No. ITMS26220120005 (activity 3.2.).

\end{document}